\documentstyle[aps,preprint,psfig]{revtex}

\begin{document}

\title{Pairing and persistent currents - the role of the far levels}

\author{M. Schechter}

\address{The Racah Institute of Physics, The Hebrew University, Jerusalem 
91904, Israel}


\author{Y. Imry, Y. Levinson and Y. Oreg}

\address{Department of Condensed Matter Physics, The Weizmann Institute of 
Science, Rehovot 76100, Israel}

\maketitle

\begin{abstract} 

We calculate the orbital magnetic response to Aharonov Bohm flux 
of disordered metallic rings with attractive pairing interaction. 
We consider the reduced BCS model, and obtain the result as an
expansion of its exact solution to first order in the interaction. 
We emphasize the connection between the large magnetic response and the 
finite occupation of high energy levels in the many-body ground state of 
the ring.

\end{abstract}

\section{Introduction}

One of the remarkable phenomena of mesoscopic physics is the existence
of equilibrium persistent currents in small normal metal rings, in the
regime where the elastic mean free path $l$ is much smaller than the
ring's circumference $L$.  This was both predicted
theoretically\cite{BIL83} and observed 
experimentally.\cite{LDDB90,CWB+91,MCB93,RRBM95,JMKW01,DBR+01} 
For the ensemble-averaged persistent current, the experimental 
results\cite{LDDB90,RRBM95,JMKW01,DBR+01} are much larger than the value 
obtained using the model of noninteracting electrons,\cite{OR91,AGI91} and 
show predominantly diamagnetic response at zero flux. In an attempt 
to resolve this discrepancy, the contribution of electron-electron 
(``e-e'') interactions was calculated\cite{AE90} and it was later suggested 
that the diamagnetic response is due to effective attractive e-e 
interactions.\cite{AE90b} Though indeed the inclusion of e-e interactions 
increased the theoretical value, it was still smaller by a factor of 
about 5 
than the experimental results, for both repulsive and attractive 
interactions.\cite{AE90b} 

In a recent work\cite{SIOL02} 
we suggested that the contribution of high energy levels 
(denoted ``far levels''), 
further than the Thouless energy $E_{\rm Th}$ from the Fermi energy 
$E_{\rm F}$, results in an enhanced orbital magnetic response (the derivative 
of the persistent current at zero flux). This was shown for the model of 
attractive pairing interactions described by the reduced BCS Hamiltonian, 
by doing perturbation theory in both the magnetic field and the e-e 
interaction. 
In this paper we 
repeat the calculation for the same model, doing 
perturbation theory in the magnetic field only, using the exact many-body 
states of the system with e-e interactions. 
In this way we can treat on the same footing different regimes of 
the strength of the interaction.  
We consider briefly the limits of zero interaction and the 
opposite limit where the interaction is strong enough and the system 
is superconducting. We then treat in more detail the case of weak e-e 
interaction, using
Richardson's exact solution for the reduced BCS Hamiltonian.\cite{Ric63,RS64} 
In this way we obtain [see Eq.~(\ref{didphiresult})] 
the result of Ref.~[\ref{SOIL02}] 
as a leading order expansion of an exact solution, instead 
of first order perturbation theory in the interaction. Furthermore, the 
present derivation has the merit of emphasizing the connection between the 
enhanced magnetic response and the pairing correlations of {\it all} the 
levels up to the high energy cutoff at the Debye frequency $\omega_{\rm D}$. 
These correlations exist in the exact many-body ground state (g.s.) 
of the ring 
[see Eqs.~(\ref{Rgs}),(\ref{Rgsf}) and discussion after Eq.~(\ref{gsimat})]. 
The contribution of the pairing correlations of the far 
levels was first realized in connection to
superconductivity in small grains.\cite{SILD01} There 
it was shown that this
contribution results in a much larger condensation energy than that
given by the BCS theory in a large regime in which
superconducting correlations are well developed, as well as in a correction 
to the spin magnetization and susceptibility as function of magnetic field 
$H$ that persists up to $H=\omega_{\rm D}/\mu_{\rm B}$.

\section{Magnetic response}

We consider a quasi one dimensional disordered ring with a finite number 
of electrons $N_{\rm e}$, 
penetrated by a constant Aharonov Bohm (AB) 
flux in its middle. We take the case of zero temperature, and the 
average spacing between noninteracting energy levels is $d$. 
The Hamiltonian is given by 

\begin{equation}
H = \sum_{\alpha} \int dr \, \psi^{\dagger}_{\alpha} (r) 
\left[\frac{1}{2m}(\vec{P} - \frac{e}{c}\vec{A})^2 + 
U(r)\right] \psi_{\alpha} (r) + H_{\rm ee} 
\label{generalH}
\end{equation}
where $U(r)$ is the external potential which includes the disorder,
and $H_{\rm ee}$ represents the e-e interactions. The vector
potential corresponding to the AB flux $\Phi$ in the middle of the ring is
given in the London gauge by $\vec{A} = \Phi/(2 \pi \rho) \hat{\phi}$, 
where $\rho$ is the distance from the origin and the angle 
$\hat{\phi}$ is in the clockwise direction of the ring. 
The ground state energy of the system is flux dependent, and can be written 
for small flux as 

\begin{equation}
E(\Phi) = E_0 - \frac{1}{2} E_2 \Phi^2 + ...
\label{EPhi}
\end{equation} 
The persistent current is given by $I = -dE/d\Phi$.\cite{Imr02} Since, due to 
time reversal symmetry 
there is no linear term of the energy 
as function of flux, hence $I(0) = 0$ and 
$dI/d\Phi\mid_{\Phi = 0} = E_2$. 

Since we are interested in $E_2$, we calculate the ground state energy 
of the system to second order in the flux. 
We take as the unperturbed Hamiltonian 

\begin{equation}
H_0 = \sum_{\alpha} \int dr \, \psi^{\dagger}_{\alpha} (r) 
\left[\frac{P^2}{2m} + U(r) \right] \psi_{\alpha} (r) + H_{\rm ee} 
\label{Hzero}
\end{equation}
and the magnetic field represented by the vector potential 
as perturbation 

\begin{equation}
H_I = \sum_{\alpha} \int dr \, \psi^{\dagger}_{\alpha} (r) 
\left[ - \frac{e}{2 m c}(\vec{P} \cdot \vec{A} + \vec{A} \cdot \vec{P}) + 
\frac{e^2}{2 m c^2} A^2 \right] \psi_{\alpha} (r) \, .  
\label{HI}
\end{equation}

We assume that the width of the ring is much smaller than its 
circumference $L$.\cite{fluxfield}
Then the $A^2$ term gives the ``diamagnetic'' contribution, 

\begin{equation}
\frac{1}{2} \left(\frac{e \Phi}{m c L}\right)^2 \hat{N} \; ,
\label{diaterm}
\end{equation} 
which is independent of the e-e interactions and results in 

\begin{equation}
E^{\rm dia}_2 = - \left(\frac{e \Phi}{m c L}\right)^2 N_{\rm e} \; .
\label{diaE2}
\end{equation} 

We denote by $| i \rangle$ the eigenstates of the noninteracting electrons in 
the disordered ring without magnetic field. 
In this basis the first and relevant term of $H_I$ is 

\begin{equation}
H^1_{\rm I} = - \sum_{i j \alpha}\frac{e \Phi}{m c L} P_{ij} 
c^\dagger_{i \alpha} c^{}_{j \alpha} \; . 
\label{HIb}
\end{equation} 
Here $c_i$ destroys an electron in the state $| i \rangle$ with 
wavefunction $\chi_i (r)$ and $P_{ij}=\langle 
i |P_{\|}| j \rangle$ is the matrix element of the momentum 
parallel to the ring's direction. 
We choose the $\chi_i$'s to be real, and then $P_{ij}$ is pure
imaginary and $P_{ii}=0$.  
Using second order perturbation theory in $H^1_I$ we write the
paramagnetic part of $E_2$ as

\begin{equation}
E^{\rm par}_2 = -2 \left(\frac{e}{mcL}\right)^2 
\sum_I \sum_{ijkl,\alpha \alpha'} 
\frac{\left\langle {\rm g.s.} | P_{ij} c^\dagger_{i \alpha} 
c^{}_{j \alpha} | I 
\right\rangle \left\langle I | P_{kl} c^\dagger_{k \alpha'} c^{}_{l \alpha'} | 
{\rm g.s.} \right\rangle}{E_{\rm g.s.} - E_I}
\label{E2per}
\end{equation}
where $E_I$ are the energies of the intermediate states $|I \rangle$. 
We model the e-e interactions by the reduced BCS interaction 

\begin{equation} 
H_{\rm ee} = -\lambda d \sum_{ij} \, \hspace{-0.1cm} ' \, 
c^\dagger_{i \uparrow} 
c^\dagger_{i \downarrow} c^{}_{j \downarrow} c^{}_{j \uparrow} \; , 
\label{HBCS}
\end{equation} 
where $\lambda$ is the dimensionless pairing parameter and the sum is over 
all levels with energies between $E_{\rm F} - \omega_{\rm D}$ and 
$E_{\rm F} + \omega_{\rm D}$. 
This interaction Hamiltonian is the usual one used when 
discussing superconducting grains, both in the perturbative and 
nonperturbative regimes\cite{DR01}
and its validity is discussed in, e.g. 
Refs.~[\ref{DR01},\ref{AA97},\ref{Aga99}]. (In 
particular, for the model to be valid the grain's dimensionless conductance g 
must be much larger than one.) 
In this model the ground state has no singly occupied 
noninteracting states\cite{Ric63,RS64} (we assume, for simplicity, that the 
number of electrons in the ring is even). 
Therefore, $|I \rangle$ has two singly occupied states, with 
opposite spins. We denote by $I_{mn}$ the set of many-body states where 
state $m(n)$ in occupied with one electron with spin up (down).
Eq.~(\ref{E2per}) can then be written as 

\begin{equation}
E^{\rm par}_2 = -2 \left(\frac{e}{mcL}\right)^2 \sum_{mn} 
\sum_{I \in I_{mn}} |P_{mn}|^2 
\frac{\left| \left\langle I | c^\dagger_{m \uparrow} c^{}_{n \uparrow} -  
c^\dagger_{n \downarrow} c^{}_{m \downarrow} 
| {\rm g.s.} \right\rangle \right|^2}{E_{\rm g.s.} - E_I} \; .
\label{E2permn}
\end{equation}

We now analyze this equation for the cases of $\lambda=0$ (normal metal), 
$\lambda > 1/\ln{N}$ 
(superconductor, see Ref.~[\ref{SILD01}]), and 
$0 < \lambda \ll 1/\ln{N}$ (weak attractive interactions). 
Here $N \equiv \omega_{\rm D}/d$. 

For $\lambda = 0$ the ground state is the noninteracting Fermi state, and 
the only relevant intermediate states are those with one electron-hole pair 
(the lowest energy states within each subspace $I_{mn}$). A straightforward 
calculation results in 

\begin{equation} 
E^{\rm par(n)}_2 = \left( \frac{2 e}{m c L} \right)^2 \sum_{m>0,n<0} 
\frac{\left|P_{mn} \right|^2}{\omega_{mn}}
\label{E2n}
\end{equation}
where $m>0$ denotes states with energies larger than $E_{\rm F}$ and 
$\omega_{mn} = \epsilon_m - \epsilon_n$, the difference between the energies 
of the single particle states $m$ and $n$. 
For a diffusive ring 
this paramagnetic term is of the same order as the diamagnetic term in 
Eq.~(\ref{diaE2}). The difference between these terms is of the order 
of the contribution of the last level (and can therefore have either sign) 
and constitutes the noninteracting sample specific result for the magnetic 
response.\cite{Imr02} 
 
The opposite limit is the superconducting regime. In this regime one can 
use the BCS approximation for the ground and excited states of the system, 
and the Bogoliubov transformation for the creation and annihilation 
operators. Eq.~(\ref{E2permn}) is then reduced to

\begin{equation} 
E^{\rm par(BCS)}_2 = 2 \left( \frac{e}{m c L} \right)^2 \sum_{m,n} 
\frac{\left|P_{mn} \right|^2 (u_m v_n - u_n v_m)^2}{E_m + E_n} \; .
\label{E2BCS}
\end{equation}
Here $u_m, v_m$ are the coherence factors, and $E_m$ is the energy of the 
electron-hole quasiparticle of state $m$. In the ballistic case, 
$P_{mn} = P_m \delta_{mn}$ and therefore the paramagnetic term is zero. 
This results in the well known perfect diamagnetism of a superconductor. 
For the diffusive case the ensemble-averaged 
momentum matrix elements are given by\cite{SIOL02} 

\begin{equation}
\left\langle |P_{mn}|^2 \right\rangle = 
\frac{p_{\rm F}^2 \tau d}{\pi 
(1 + \omega_{mn}^2 \tau^2) s}
\label{pmnaverage}
\end{equation}
which is roughly constant for $\omega_{mn} < 1/\tau$ and zero for 
$\omega_{mn} > 1/\tau$ ($\tau$ is the elastic mean free time and 
$s=1,2,3$ is the effective dimension of the ring for diffusive motion). 
The total response in this case (sum of diamagnetic and paramagnetic terms) 
is diamagnetic, with an approximate magnitude of $(l/\xi) E^{\rm dia}_2$, 
where $\xi$ is the (ballistic) superconducting coherence length. 

We now turn to the calculation of $E^{\rm par}_2$ for the case of weak 
attractive interaction ($\lambda < 1/\ln{N}$). 
We calculate the interaction correction to $E_2$ to first order in 
$\lambda$. Using Richardson's exact solution\cite{RS64} one finds that 
to first order in $\lambda$ the ground state can be written as 

\begin{equation}
\Psi_{\rm g.s.} = \sum_{\{f_1...f_N\}} \phi(f_1...f_N) b^\dagger_{f_N} ...
b^\dagger_{f_1} | vac \rangle
\label{Rgs}
\end{equation}
with 
\begin{eqnarray}
\phi(1...N) & = & 1 \nonumber \\
\phi(1...N;\neq j,k) & = & \frac{\lambda d}{2(\epsilon_k - \epsilon_j)} \; .
\label{Rgsf}
\end{eqnarray}
Here $(f_1...f_N)$ denotes a set of $N$ out of the $2N$ noninteracting 
eigenstates between $E_{\rm F} - \omega_{\rm D}$ and 
$E_{\rm F} + \omega_{\rm D}$, $b^\dagger_{f_1} \equiv 
c^\dagger_{f_1 \uparrow} c^\dagger_{f_1 \downarrow}$ and  
$\phi(1...N;\neq j,k)$ is the amplitude of the many-body state with 
a filled Fermi sea except a pair excitation from state $j$ below 
$E_{\rm F}$ to state $k$ above $E_{\rm F}$. 
To first order in $\lambda$, the amplitude of all the other possible 
configurations is zero for the ground state. 
The finite amplitude to occupy noninteracting states with energies larger 
than $E_{\rm F}$ is a result of the interaction. For any small $\lambda$ the 
system gains energy by having a different ground state than the Fermi state. 
Note that the amplitudes of all the different configurations come with the 
{\it same} sign. 
Similar analysis of the intermediate states shows that the only intermediate 
states that contribute to $E_2$ in first order in $\lambda$ are 
$I^{\rm l}_{mn}$, the lowest 
energy states of each subspace $I_{mn}$, with $m$ and $n$ on opposite sides 
and within $\omega_{\rm D}$ of the Fermi surface. Then 

\begin{equation}
\left\langle I^{\rm l}_{mn} | c^\dagger_{m \uparrow} c^{}_{n \uparrow} - 
c^\dagger_{n \downarrow} c^{}_{m \downarrow} | g.s. \right\rangle = 1 - 
\frac{\lambda d}{2 \omega_{mn}}
\label{gsimat}
\end{equation}
and $E_{\rm g.s.} - E_I = \omega_{mn} + \lambda d$. 
Thus, the pairing interaction contributes to the magnetic response in two 
ways. First, due to the finite occupancy of levels above $E_{\rm F}$ the 
matrix element in Eq.~({\ref{gsimat}) has a contribution not only from 
the annihilation of an electron in state $n$ below $E_{\rm F}$ and the 
creation of an electron in state $m$ above $E_{\rm F}$. Since in the 
ground state there is a finite amplitude for state $m$ to be doubly 
occupied and state $n$ to be empty, there is a finite contribution to the 
matrix element from annihilating and electron in state $m$ above $E_{\rm F}$
and creating it in state $n$ below $E_{\rm F}$ [second term in 
Eq.~({\ref{gsimat})].
Second, it adds a term $\lambda d$ to the energy of the 
excited states due to the excess 
pairing energy of a doubly occupied state. The latter term is a Hartree-like 
contribution, coming from the diagonal part of the pairing interaction, 
while the former is due to pairing correlations of pairs in 
different single-particle states [offdiagonal part of the Hamiltonian in 
Eq.~(\ref{HBCS})]. 
Both of these contributions suppress the paramagnetic term in the case 
of attractive interaction, and as a result, to first order in $\lambda$, 

\begin{equation}
E^{\rm par(fo)}_2 = E^{\rm par(n)}_2 - \lambda d 
\left(\frac{2 e}{m c L}\right)^2 \sum_{m>0,n<0} 
\frac{\left|P_{mn}\right|^2}{\omega_{mn}^2} \; .
\label{Eparfo}
\end{equation}
Using Eq.~(\ref{pmnaverage}) we find that 
the derivative of the persistent current at zero flux, to first order in 
the interaction, is given by

\begin{equation}
\left\langle E^{\rm fo}_2 \right\rangle 
= \frac{8 \pi \lambda E_{\rm Th}}{\Phi_0^2} 
\ln{\frac{\omega_{\rm D}}{d}} \; .
\label{didphiresult}
\end{equation} 
In comparison with the known result for the first order interaction 
correction to the magnetic response\cite{AE90}, 
our result has a much larger logarithmic cutoff, 
$\omega_{\rm D}$ compared with $E_{\rm Th}$. This enhancement is irrespective 
of the higher order correction, which for attractive interaction further 
increases the first order result.\cite{AE90b} The large logarithm
we obtain is a result of enhanced pairing correlations of all the states 
within $1/\tau$ from $E_{\rm F}$. We expect that semiclassically this term 
necessitates only one circulation of the ring, and will therefore affect 
the magnitude of the persistent current as well 
(see discussion in Ref.~[\ref{SOIL02}]). This may lead to a resolution 
of the discrepancy between the experimental results and theory of persistent 
currents in diffusive normal rings.

\section*{Acknowledgments}

We benefited from valuable discussions with N. Argaman, A. Punnoose,
B. Altshuler, H. Bouchiat, J. von Delft, A. Finkel'stein, Y. Gefen,
D. Gobert, M. Khodas, D. Orgad, Z. Ovadyahu, U. Smilansky, and
R. A. Webb.  M.S. is thankful for the support by the Lady Davis fund.
This work was supported by a Center of Excellence of the Israel
Science Foundation, Jerusalem and by the German Federal Ministry of
Education and Research (BMBF) within the Framework of the
German-Israeli Project Cooperation (DIP) and by the German-Israeli
Foundation(GIF).


\begin{thebibliography}{99}
\bibitem{fluxfield}
Under this assumption of thin ring the response to constant magnetic field can
  be found along the same lines as done here for AB flux.
\end{thebibliography}
\end{document}